\documentstyle[preprint,aps]{revtex}
\tighten

\expandafter\chardef\csname pre amssym.def at\endcsname=\the\catcode`\@
\catcode`\@=11

\def\undefine#1{\let#1\undefined}
\def\newsymbol#1#2#3#4#5{\let\next@\relax
 \ifnum#2=\@ne\let\next@\msafam@\else
 \ifnum#2=\tw@\let\next@\msbfam@\fi\fi
 \mathchardef#1="#3\next@#4#5}
\def\mathhexbox@#1#2#3{\relax
 \ifmmode\mathpalette{}{\m@th\mathchar"#1#2#3}%
 \else\leavevmode\hbox{$\m@th\mathchar"#1#2#3$}\fi}
\def\hexnumber@#1{\ifcase#1 0\or 1\or 2\or 3\or 4\or 5\or 6\or 7\or 8\or
 9\or A\or B\or C\or D\or E\or F\fi}

\font\tenmsa=msam10
\font\sevenmsa=msam7
\font\fivemsa=msam5
\newfam\msafam
\textfont\msafam=\tenmsa
\scriptfont\msafam=\sevenmsa
\scriptscriptfont\msafam=\fivemsa
\edef\msafam@{\hexnumber@\msafam}
\mathchardef\dabar@"0\msafam@39
\def\dashrightarrow{\mathrel{\dabar@\dabar@\mathchar"0\msafam@4B}}
\def\dashleftarrow{\mathrel{\mathchar"0\msafam@4C\dabar@\dabar@}}

\def\ulcorner{\delimiter"4\msafam@70\msafam@70 }
\def\urcorner{\delimiter"5\msafam@71\msafam@71 }
\def\llcorner{\delimiter"4\msafam@78\msafam@78 }
\def\lrcorner{\delimiter"5\msafam@79\msafam@79 }
\def\yen{{\mathhexbox@\msafam@55 }}
\def\checkmark{{\mathhexbox@\msafam@58 }}
\def\circledR{{\mathhexbox@\msafam@72 }}
\def\maltese{{\mathhexbox@\msafam@7A }}

\font\tenmsb=msbm10
\font\sevenmsb=msbm7
\font\fivemsb=msbm5
\newfam\msbfam
\textfont\msbfam=\tenmsb
\scriptfont\msbfam=\sevenmsb
\scriptscriptfont\msbfam=\fivemsb
\edef\msbfam@{\hexnumber@\msbfam}

\def\widehat#1{\setbox\z@\hbox{$\m@th#1$}%
 \ifdim\wd\z@>\tw@ em\mathaccent"0\msbfam@5B{#1}%
 \else\mathaccent"0362{#1}\fi}
\def\widetilde#1{\setbox\z@\hbox{$\m@th#1$}%
 \ifdim\wd\z@>\tw@ em\mathaccent"0\msbfam@5D{#1}%
 \else\mathaccent"0365{#1}\fi}
\font\teneufm=eufm10
\font\seveneufm=eufm7
\font\fiveeufm=eufm5
\newfam\eufmfam
\textfont\eufmfam=\teneufm
\scriptfont\eufmfam=\seveneufm
\scriptscriptfont\eufmfam=\fiveeufm

\catcode`\@=\csname pre amssym.def at\endcsname

\newsymbol\lessapprox 132F

\begin{document}
\preprint{CERN-TH.7500/94/Revised {\hspace{7.5 cm}} astro-ph/9507051}


\draft


\title{Experimental limits to the
density of dark matter\\ in the solar system}

\author{{\O}yvind Gr{\o}n}

\address{Oslo College, Faculty of Engineering, Cort Adelers Gate 30, N-0254
Oslo, Norway}
\address{Institute of Physics, University of Oslo,
         P.O. Box 1048, N-0316 Blindern, Oslo 3, Norway}

\author{Harald H. Soleng}

\address{CERN, Theory Division,  CH-1211 Geneva 23, Switzerland}

\date{16 November 1994; revised 8 June 1995}

\bibliographystyle{unsrt}

\maketitle


\begin{abstract}

\noindent
On the scales of galaxies
and beyond there is
evidence for unseen
dark matter.
In this paper we find the experimental limits to
the density of dark matter bound
in the solar system by studying its
effect upon planetary motion.
\end{abstract}

\pacs{{\mbox{ }}\\
Subject Headings: Gravitation --- Dark Matter (solar system tests)
{\mbox{ }}\\
{\mbox{ }}\\
{\mbox{ }}\\
{\mbox{ }}\\
{\mbox{ }}\\
{\mbox{ }}\\
{\mbox{ }}\\
{\mbox{ }}\\
{\mbox{ }}\\
{\mbox{ }}\\
{\mbox{ }}\\
{\mbox{ }}\\
{\mbox{ }}\\
{\mbox{ }}\\
To appear in the Astrophysical Journal}

%
%
\renewcommand{\thesection}{\arabic{section}}

\section{Introduction}

According to Newton's inverse square force law, the circular speed
around an
isolated object of mass $M$ should be
\[
v_{c}=\sqrt{\frac{MG}{r}}.
\]
In disk galaxies we do, however, observe that the circular speeds
are approximately independent of $r$
at large distances.
The standard explanation is that this is due to
halos of unseen matter that makes up
around 90\% of the total mass of the galaxies
(Tremaine 1992).
The same pattern repeats itself on larger and larger scales,
until we reach the cosmic scales where a baryonic
density compatible with successful big bang nucleosynthesis
is less than 10\% of the density predicted by inflation, i.e.\
the critical density.

The flat rotation curves of galaxies, taken at face value,
imply that
the effective gravitational force follows
a $1/r$ law at large scales.
This could either be due to dark matter or to a
departure from Newtonian dynamics at small
accelerations (Milgrom 1983; Bekenstein 1992) 
or large scales (Sanders 1990). 
An effective gravitational acceleration
law of the form
\[
g=- \frac{\sqrt{GM a_{0}}}{r}
\]
at small accelerations $a\ll a_{0}$ has been reported
(Kent 1987; Milgrom 1988; Begeman, Broeils, \& Sanders 1991)\
to be
successful
in reproducing the observations of galactic
systems.\footnote{However, not without debate
(Lake 1989; Milgrom 1991).}
The constant $a_{0}$ has been determined by studies
of galaxy rotation curves
and its value has been found to be $a_{0}\approx 10^{-8}$~cm~s$^{-2}$.
As noted by Milgrom (1983),
this value of
$a_{0}\approx c H_{0}$.

With such a $1/r$ force law the circular speed would approach
$v_{c}= (GM a_{0})^{1/4}$.
If the luminosity $L$ of a galaxy is proportional to its mass $M$, then
this relation would explain the
infrared Tully--Fisher law (Tully \& Fisher 1977)\ 
which states that circular speeds in galaxies scale
as
$v_{c}\propto L^{1/4}$.

The theoretical underpinning for the $1/r$
effective force law is not yet firmly established.
It might
be due to a modification of gravity
along the lines of Milgrom (1983),  
but it seems to be difficult to construct a viable relativistic theory
of this kind (Zhytnikov and Nester, 1994).
Accordingly, the standard view is that
the effective galactic $1/r$ force law is caused by dark matter.
At this point, it is worth mentioning that
a large-distance force law of this
type can be reproduced
within
standard general relativity theory with a very simple,
but perhaps unrealistic,  matter source
(Soleng 1993, 1995).
Our key point is that general relativity
is quite capable of explaining the observed
gravitational properties of the
universe provided we give it the right input.
Most likely the dark matter is a mixture
of several components, such as weakly interacting particles, black holes,
brown dwarfs,
neutron stars, as well as
energy stored in high-frequency oscillation of Newton's
gravitational coupling
(Accetta \& Steinhardt 1991; Steinhardt \& Will 1994).
Whatever the origin of the $1/r$ force law might be,
its
reported experimental
success forces us to take it seriously.
Accordingly, we think that
it is particularly important to compare the densities
of dark matter inferred from large scale dynamics with
experimental limits from local tests. If dark matter exists
in the
form of {\em microscopic\/} objects, one would expect that this unknown
form of energy
penetrates into galaxies and also enter the solar system.

Braginsky, Gurevich, and Zybin (1992)\ 
have studied the effect of dark matter bound in the galaxy but unbound to the
solar system. Such unbound dark matter would produce an anisotropy in the
gravitational background of the solar system. The resulting tidal forces induce
an additional perihelion precession.
Assuming $\rho_{d}=0.3$~GeV/cm$^3=5.4 \times 10^{-25}$~g/cm$^{3}$
they computed the magnitude of the resulting
secular orbit distortion.
The
effect may be observed by reasonable improvements of
present observational techniques
(Klioner and Soffel 1993; Braginsky 1994).
A possible influence of dark matter on the Earth--Moon system has
been considered by Nordtvedt (1994)\ and by Nordtvedt, M\"uller,
and Soffel (1995).

In this paper we focus on
a dark matter
model in which
the density of dark matter varies
so slowly within the solar system that it can considered
constant.
This is a reasonable assumption if dark matter
in the solar system really is
in the much deeper potential of the galaxy
with the Sun causing only a local density perturbation
in the galactic dark matter background.
It will also be assumed that
the equation of state of
the unseen matter is almost dust-like,
that is, the pressure will
be assumed to be much less than the energy-density.
Based on
this model we
calculate an upper limit to the
density of dark matter by considering its effect upon the perihelion
precession of the planets.
We have also carried out similar computations with a dark matter
density proportional to $1/r^4$ and $1/r^2$ using the results in
Soleng (1994) and (1995), respectively. The corresponding
experimental bounds do not vary more than one order of magnitude.
This weak dependence on the distribution
function
corroborates the claim of
Anderson et al.\ (1989) and
should be expected because
(to lowest order) the perihelion precession caused by dark matter
is given by
the integrated dark matter mass at a given orbital radius.

\section{Solar system with dark matter}

In order to study the gravitational effects of
hypothetical dark matter on planetary motion,
we need a solution of Einstein's field equations for
a static, spherically symmetric space--time
and a given
distribution of
dark matter.
The line-element for a static, spherically symmetric gravitational field
can in general be written
as\footnote{We employ
geometrized units with $G=c=1$.}
\begin{equation}
ds^2=
-e^{2\mu(r)}dt^2
+e^{2\lambda(r)} dr^2
+r^2 d\Omega^2.
\label{metric}
\end{equation}
We shall assume that the dark matter has a
constant density $\rho_{0}$
(within the solar system).
At a surface where the dark matter pressure equals the
galactic dark matter pressure
$p_{\text{G}}\approx 10^{-7} \rho_{\text{G}}$ (characterized by velocities of
$220 {\mbox{km}}/{\mbox{s}}$),
we match the gravitational field of the solar system to
the exterior
field of the galaxy.
We shall assume that $g_{tt}\approx -1$ at this
distance (this assumption is always used in
local gravitational problems).
Then the $tt$-component of the field
equations takes the form
\[
\frac{d}{dr}\left[ r \left(1-e^{-2\lambda}\right)\right]=8\pi \rho_{0} r^2.
\]
Integrating with a spherical mass $M$ (the Sun) at $r=0$
gives
\[
e^{-2\lambda}=1-\frac{2 M}{r}-\frac{8\pi}{3}\rho_{0} r^2.
\]
The Tolman--Oppenheimer--Volkov (TOV) equation takes the form
\begin{equation}
\frac{dp}{dr}=-(\rho_{0}+p)\frac{M+\frac{4\pi}{3}\rho_{0}r^3+4\pi p r^3}{
r^2-2 r (M+\frac{4\pi}{3}\rho_{0}r^3)},
\label{TOV}
\end{equation}
where $p$ is the pressure of the dark matter.
Using that the mass of dark matter
in the solar system is much less than the mass of the Sun,
we can neglect the last term in the denominator
of equation~(\ref{TOV}).
Also we assume that the dark matter
is non-relativistic, {\em i.e.}\ that $p\ll\rho_{0}$,
and accordingly we can neglect the last term in the numerator.
Then the TOV equation reduces to
\[
\frac{dp}{dr}=-(\rho_{0}+p)\frac{M+\frac{4\pi}{3}\rho_{0}r^3}{
r (r-2  M)}.
\]
Integration leads to
\[
p= \frac{K_{1}}{\sqrt{1-\frac{2 M}{r}}}
\left(\frac{r}{2 M}-1\right)^{-16 M^2 \pi \rho_{0}/3}
\exp\left[-\frac{2\pi}{3}\rho_{0}(4 M r +r^2) \right]
-\rho_{0},
\]
where $K_{1}$ is a constant of integration.
With the assumptions that
$M^2 \rho_{0}\ll 1$ and $M\ll r$,
this equation simplifies to
\begin{mathletters}
\begin{equation}
p= \frac{K_{1}}{\sqrt{1-\frac{2 M}{r}}}
e^{-\frac{2\pi}{3}\rho_{0}r^2}
-\rho_{0}.
\label{pressure}
\end{equation}
The constant $K_{1}$ can be determined as follows:
let $p_{0}$ be the pressure at $r=r_{\text{surf}}$ where
$r_{\text{surf}}$ is the surface of the central
mass (the Sun in our case).
Within our approximations,
\begin{equation}
K_{1}=\rho_{0}+p_{0}.
\label{K1}
\end{equation}%
\label{TOVint}%
\end{mathletters}%
The relativistic equation of hydrostatic equilibrium,
${T^{\mu\nu}}_{;\nu}=0$,
where $T^{\mu\nu}$ is the energy--momentum tensor of
the dark matter, leads to
\[
\frac{dp}{dr}=-(\rho_{0}+p)\frac{d\mu}{dr}.
\]
Integration yields
\[
g_{tt}=-e^{2\mu}=-K_{2} \left(\frac{\rho_{0}+p_{0}}{\rho_{0}+p}\right)^2
\]
where $K_{2}$ is a new integration constant.
Inserting the pressure from equations~(\ref{TOVint})
leads to
\[
g_{tt}=-K_{2} \left(1-\frac{2 M}{r}\right)
e^{\frac{4 \pi}{3} \rho_{0} r^2}.
\]
The constant $K_{2}$ can be determined
by demanding that $g_{tt}=-1$
at the surface $r=r_{\text{match}}$ where the
pressure equals the galactic dark matter pressure
$p_{\text{G}}$.
This radius is given by
\begin{equation}
r_{\text{match}}=
\left[\frac{3}{2\pi\rho_0} \ln
\left(\frac{\rho_0+p_0}{\rho_0+
p_{\text{G}} }\right)
\right]^{1/2}
\label{match}
\end{equation}
and hence
\[
K_{2} =
\exp\left(-\frac{4 \pi}{3} \rho_{0} r_{\text{match}}^2 \right)
=
\left(\frac{\rho_0+p_{\text{G}}}{\rho_0+
p_0 }\right)^2
\approx 1
\]
according to the assumption that $p\ll\rho_{0}$.

Our model of the dark matter filled
space--time
in the solar system is thus
represented by the approximate line-element
\begin{equation}
ds^2=
-\left(1-\frac{2 M}{r}
+\frac{4 \pi}{3} \rho_{0} r^2
\right)
dt^2+
\frac{dr^2}{1-\frac{2 M}{r} - \frac{8\pi}{3}\rho_{0} r^2}
+ r^2 d\Omega^2
\label{line-element}
\end{equation}
valid outside the Sun and for $r \ll (M/\rho_{0})^{1/3}$.
It should be noted that we only require the validity of this
expression inside the solar system, where
$\frac{4\pi}{3}\rho_{0} r^2$ is a small number.
At larger scales where the dark matter pressure equals its
galactic value, the gravitational field
is of course
determined by the mass distribution
of the galaxy.

\section{Perihelion precession}

The  Lagrange function
for a test particle moving in the $\theta=\pi/2$ plane
in the geometry specified by
equation (\ref{line-element}),  is
\[
2 L = -
\left(1-\frac{2 M}{r}
+
\frac{4 \pi}{3} \rho_{0} r^2
\right)
\dot{t}^2 +
\frac{\dot{r}^2}{
1-\frac{2 M}{r} - \frac{8\pi}{3}\rho_{0} r^2}
+ r^2 \dot\phi^2
\]
where
a dot denotes differentiation with respect to the proper time $\tau$
of the particle (a planet).
The momenta $p_{\mu}\equiv \partial L/\partial \dot x^{\mu}$
are
\begin{mathletters}
\begin{eqnarray}
p_{t}&=&
\left(1-\frac{2 M}{r}
+
\frac{4 \pi}{3} \rho_{0} r^2
\right)
\dot t = {\mbox{constant}}
\label{Pt}\\
p_{r}&=&
\left( 1-\frac{2 M}{r} - \frac{8\pi}{3}\rho_{0} r^2 \right)^{-1}
\dot r
\label{Pr}\\
p_{\phi} &=& r^2\dot\phi = {\mbox{constant}}.
\label{Pphi}
\end{eqnarray}
\end{mathletters}%
Using the normalization of the momenta
$g_{\mu\nu}p^{\mu}p^{\nu}=-1$,
we get
\[
\frac{p_{t}^2}{g_{tt}}+
\frac{p_{r}^2}{g_{rr}}+
\frac{p_{\phi}^2}{g_{\phi\phi}}
=
-1.
\]
To first order in $M$ and $\rho_{0}$, one finds
\begin{equation}
\dot r^2=
\left(
1-\frac{2 M}{r} - \frac{8\pi}{3}\rho_{0} r^2 \right)
\left(
-1-\frac{p_{\phi}^2}{r^2}\right) +p_{t}^2-4\pi p_{t}^2\rho_{0} r^2.
\label{rdotsquare}
\end{equation}
Inserting the new radial variable $u\equiv 1/r$
and
noting from
equation~(\ref{Pphi})\ that
$\dot r=-p_{\phi} (du/d\phi)$,
brings equation~(\ref{rdotsquare}) into the form of an orbit-equation.
Differentiation of the resulting expression, and
using that $p_{t}\approx 1$ for
planets in the solar system,
leads to
\begin{equation}
\frac{d^2 u}{d\phi^2} +u = \frac{M}{p_{\phi}^2}
+3 M u^2+\frac{4\pi}{3} \frac{\rho_{0}}{u^3 p_{\phi}^2}.
\label{orbiteq}
\end{equation}
The planetary orbits are nearly circular, and we can treat the
perihelion precession as a perturbation from the circular solution,
$u=u_{0}$, where
\begin{equation}
u_{0}=
\frac{M}{P_{\phi}^2} + 3 M u_{0}^2 +
\frac{4\pi}{3} \frac{\rho_{0}}{u_{0}^3 p_{\phi}^2}.
\label{circularsolution}
\end{equation}
Substituting $u_{0}(1+\varepsilon)$ for $u$
into equation~(\ref{orbiteq})\
with $\varepsilon\ll 1$,
and using
equation (\ref{circularsolution})\ and calculating to first order in
$\varepsilon$,
we get
\[
\frac{d^2 \varepsilon}{d\phi^2}=
\left(
 6 M u_{0} -1
-4\pi \frac{\rho_{0}}{u_{0}^4 p_{\phi}^2} \right) \varepsilon.
\]
The Einstein precession coming from the solar mass $M=M_{\odot}$
 is
\begin{equation}
\Delta\phi_{0}=
6 \pi M_{\odot} u_{0}.
\label{EinsteinPres}
\end{equation}
In addition, there is a dark matter induced precession
\begin{equation}
\Delta\phi_{\text{dark}}= -4\pi^2 \frac{\rho_{0}}{u_{0}^4 p_{\phi}^2}.
\label{DarkPres}
\end{equation}
Equations~(\ref{EinsteinPres}) and (\ref{DarkPres}) imply
\begin{equation}
\Delta\phi_{\text{dark}}= \frac{2\pi}{3}
\frac{\rho_{0}}{M_{\odot} u_{0}^5 p_{\phi}^2}
\Delta\phi_{0}.
\label{PR}
\end{equation}
Let the observed
non-Newtonian
perihelion precession
be denoted
by
$\Delta\phi_{\text{obs}}$, and its uncertainty
by $\delta\phi_{\text{obs}}$.
Since $\Delta\phi_{\text{obs}}=\Delta\phi_{0}$ within the
uncertainty,
the dark matter perihelion precession
and the Einstein term
are related by
$|\Delta\phi_{\text{dark}}|
\leq
|\delta\phi_{\text{obs}}|$.
This, together with equation~(\ref{PR}),
gives
\[
\rho_{0} \lessapprox \frac{3}{2\pi} M_{\odot} p_{\phi}^2 u_{0}^5
\frac{|\delta \phi_{\text{obs}}|}{\Delta\phi_{0}}.
\]
To lowest order $p_{\phi}^2=M/u_{0}$. Thus
\[
\rho_{0} \lessapprox \frac{3}{2\pi} \frac{G M_{\odot}^2}{c^2 r_{0}^4}
\frac{|\delta \phi_{\text{obs}}|}{\Delta\phi_{0}}.
\]
where we have inserted Newton's constant $G$ and the speed
of light $c$
in order to simplify the numerical
calculations.
The perihelion precession of the asteroid Icarus is known with about
8~\%\ accuracy. Its distance from the Sun is 1.076~A.U\@.
With
$M_{\odot}=2 \times 10^{33}$~g and $r_{0}=1.076$~A.U.$=1.61 \times
10^{13}$~cm
we obtain
\begin{equation}
\rho_{0} \lessapprox 1.8\times 10^{-16} {\mbox{ g/cm}}^3.
\label{rholimit}
\end{equation}
This value is about seven orders of magnitude above the
mean galactic mass density, and it shows that measurements of the
perihelion precession of the planets do not
put strict limits on the density of bound dark matter.
Note that with this density, the mass of spherically symmetric
bound
dark matter of constant density (\ref{rholimit}) within
the orbit of
Pluto is less than $2 \times 10^{-5}$~$M_{\odot}$
and thus in agreement with  our
assumptions.
Within Uranus we find
a limit of
\[
M_{\text{dark}}(r_{\text{U}}) \lessapprox
2 \times  10^{-6}\; M_{\odot} \approx 0.6\, M_{\oplus}
\]
where $M_{\oplus}$ is the Earth's mass.
This bound
is of the same size as the bound found by Anderson et al.\ (1989)
by numerically analyzing how the orbit
of Uranus would be affected by dark matter.
Recently, the bound obtained by such methods has been strengthened to
around $0.2\, M_{\oplus}$ (Anderseon et al.\ 1995),
but a similar improvement should
also be possible using improved perihelion precession observations.

Let us finally check the assumption that the dark matter has
a pressure which is much less than
its energy density. The maximum pressure is at the center of
the solar system. Now, if we require that the
model is valid out to the Oort cloud at several thousand
A.U. and that macthing to the
galactic dark matter distrubution takes place here,
we find that the pressure gradient necessary for hydrostatic
equilibrium is very small.
Using
equation~(\ref{match})\ and
matching at, say, $r_{\text{match}}=5000\; {\mbox{A.U.}}$
gives
\begin{equation}
\frac{p_0-p_{\text{G}}}{p_{\text{G}}}\approx 0.0015.
\end{equation}
Since $P_{\text{G}}\approx 10^{-7}\rho_0$,
this result also confirms that
$p\ll \rho_{0}$.

\section{Discussion}

Dark matter in the solar system would increase the perihelion
precession of the planets. Assuming a very simple model
for dark matter having a constant density and
being kept in hydrostatic
equilibrium by a small but positive pressure, we have used
the perihelion precession of Icarus to put a limit
of $\rho\lessapprox 1.8\times 10^{-16}\;{\mbox{g}}/{\mbox{cm}}^3$ on the
density of such dark matter in the solar system.
This bound is much larger than the mean galacitic density, but
we note that the bound is only weakly dependent on the
dark matter distribution function and that some dark matter
models may predict a local concentration of dark matter energy.

If dark matter exists in the form of ordinary matter,
one would expect that orbital perturbations, especially
by Jupiter,
effectively would clean the inner parts of the solar system
of any weakly interacting dark matter (Anderson  et al.\ 1989).
In this case, our bound which was derived
from an asteroid would be irrelevant.
However, cosmological constraints from inflationary models
and cosmic nucleosynthesis, strongly
suggest that a large part of dark matter is non-baryonic.
We do not know what the non-baryonic dark matter
is. One possibility (Accetta \&\ Steinhardt 1991;
Steinhardt \&\ Will 1994) is that the dark matter in part
is
oscillation energy caused by rapid oscillations of Newton's constant.
An effective Brans--Dicke field is a consequence of
many unification schemes and also an ingredient of extended inflationary
models
(La \&\ Steinhardt 1989).
Extended inflation would drive the scalar field away from
the minimum of its potential, and the field would
then start to oscillate when inflation ends.
This possibility  only illustrates that dark matter may
behave rather differently from normal matter and
that no stone should be left unturned in the search for
the mass which seems to make up most of the universe we inhabit.

\acknowledgements
It is a pleasure to thank Slava G. Turyshev for
pointing out some useful references. We are also
indebted to an anonymous
referee
for constructive criticism.






\newpage
\section*{References}

\noindent
Accetta, F. \& Steinhardt, P. J. 1991,
Phys.\ Rev.\ Lett.,  67, 298.

\noindent
Anderson, J. D., Lau, E. L., Taylor, A. H.,
Dicus, D. A.,
Teplitz, D. C., and Teplitz, V. L. 1989,
ApJ, 342, 539.

\noindent
Anderson, J. D., Lau, E. L.,
Krisher, T. P.
Dicus, D. A.,
Rosenbaum, D. C., and Teplitz, V. L. 1995,
ApJ (in press).

\noindent
Begeman, K. G., Broeils, A. H., \& Sanders, R. H. 1991,
MNRAS, 249, 523.

\noindent
Bekenstein, J. 1992, in Proceedings
of the 6th Marcel Grossmann Meeting on General Relativity,
ed. H. Sato \& T. Nakamura (Singapore: World Scientific Publ.),
905

\noindent
Braginsky, V. B.  1994,
Class.\ Quantum Grav., 11, A1.

\noindent
Braginsky, V. B.,  Gurevich, A. V., \&  Zybin,
K. P.
1992,
Phys.\ Lett.\ A,  171, 275.

\noindent
Kent, S. M. 1987,  AJ, 93, 816.

\noindent
Klioner, S. \&
Soffel, M. 1993, Phys.\ Lett.\ A, 184, 43.

\noindent
La, D. \&\ Steinhardt, P. J. 1989, Phys.\ Rev.\ Lett.\ 62, 376.

\noindent
Lake,
G. 1989,  ApJ,   345, L17.

\noindent
Milgrom, M. 1983, ApJ, 270, 365.

\noindent
Milgrom, M. 1988, ApJ, 333, 689.

\noindent
Milgrom,
M. 1991, ApJ,  367, 490.

\noindent
Nordtvedt, K. L. 1994,
ApJ, 437, 529.

\noindent
Nordtvedt, K. L., M\"uller, J. \&\
Soffel, M. 1995,
A\& A, 293, L7.

\noindent
Sanders, R. H. 1990,  Astron.\ Astrophys.\ Rev.,  2, 1.

\noindent
Soleng, H. H. 1993,  BAAS,   25, 796.

\noindent
Soleng, H. H. 1994,
Gen.\ Rel.\ Grav.,  26, 149.

\noindent
Soleng, H. H. 1995,
Gen.\ Rel.\ Grav., 27, 367.

\noindent
Steinhardt, P. J. \& Will, C. M. 1995, {\em High-frequency
oscillations of Newton's constant induced by Inflation,}
Phys.\ Rev.\ D (in press).

\noindent
Tremaine, S. 1992,
Physics Today, 45, 28.

\noindent
Tully,  R. B. \&
Fisher, J. R. 1977, A{\&}A,  54, 661.

\noindent
Zhytnikov, V. V. and Nester, J. M. 1994,
Phys.\ Rev.\ Lett., 73, 2950.

\end{document}